\documentclass{phb-proc4-auth}

\usepackage{graphicx}
\usepackage{amssymb}

\begin{document}
\begin{frontmatter}

%----------------------------------------------------------------------
% Specify destination and version number of the manuscript

\journal{SCES '04}

%----------------------------------------------------------------------
% Title of manuscript

\title{Coulomb blockade and Quantum Critical Points in Quantum Dots}

%----------------------------------------------------------------------
% List of authors
%
% List each author using a separate \author{} command
%
% If there is more than one author address, add a label to each author
% of the form \author[label]{name}.  This label should be identical to
% the corresponding label provided with the \address command.
%
% e.g. if there are three authors from two institutions in USA and 
% France, you can link them to their respective addresses, using
%
% \author[US]{John Doe}
% \author[US,FR]{Jane Doe}
% \author[FR]{Jean Dupont}
% \address[US]{University of Life, Somewhere, USA}
% \address[FR]{Universite de la Vie, Quelque Part, France}
%
% N.B. Unlike the document class used for abstract submissions, it is
% possible to have the author associated with more than one address,
% as shown in the example above.
%

\author[GER]{Frithjof B.~Anders}
\author[RUT]{Eran Lebanon}
\author[IS]{Avraham Schiller}

%----------------------------------------------------------------------
% List of addresses
%
% If there is more than one address, list each using a separate 
% \address command using a label to link it to the respective author
% as described above
 
\address[GER]{Department of Physics, Universit\"at Bremen,
              P.O. Box 330 440, D-28334 Bremen, Germany}
\address[RUT]{Department of Physics and Astronomy,
              Rutgers University, Piscataway, NJ 08854-8019 USA}
\address[IS]{Racah Institute of Physics, The Hebrew University,
             Jerusalem 91904, Israel}

%----------------------------------------------------------------------
% Title page footnotes
%
% If you need to add qualifying information to any of the authors, 
% use the \thanksref{} command within the \author command.  The 
% argument is the label of a corresponding \thanks[label]{text}
% command which contains the footnote text
%
% e.g. you can acknowledge a funding authority for John Doe, using
%
% \author{John Doe\thanksref{ABC}}
% \thanks[ABC]{This work was supported by Institute of Unphysical 
%    Phenomena under contract no. ABC-123}
%

%\thanks[]{}

%----------------------------------------------------------------------
% Contact Information
%
% Add the complete postal address, telephone number, fax number, and
% email address of the corresponding author as a special footnote using
% the \corauth[]{} command.  This works in a similar way to the \thanks 
% command.  Add the \corauthref{} command within the \author command.
% The argument is the label of a corresponding \corauth[label]{text}
% command which contains the contact information.  Prefix the text with
% Corresponding Author:
%
% e.g. if the contact author is John Doe,
%
% \author{John Doe\corauthref{1}}
% \corauth[1]{Corresponding Author: University of Life, 123 Some St.,
%    Somewhere, MI 12345, USA.  Phone: (555) 555-5555 
%    Fax: (555) 555-7777, Email: JDoe@uol.edu}
%

\corauth[]{}

%----------------------------------------------------------------------
% Text of abstract

\begin{abstract}
An ultrasmall quantum dot coupled to a lead and to a quantum
box (a large quantum dot) is investigated. Tuning the tunneling
amplitudes to the lead and box, we find a line of unstable
non-Fermi-liquid fixed points as function of the gate potentials
of the quantum dots, extending to arbitrary charging energies
on the small and large quantum dots. These quantum-critical
fixed points possess a finite residual entropy. They govern
the cross over from one Fermi-liquid regime to another,
characterized by distinct (high and low) conductance values.

\end{abstract}

%----------------------------------------------------------------------
% Manuscript keywords
%
% Please give two or three keywords in the form: keyword \sep keyword
% e.g. NMR \sep superconductivity
%
% NB The syntax is different from the abstract document class


%----------------------------------------------------------------------
% End of front page

\end{frontmatter}

%----------------------------------------------------------------------
% Manuscript text
%
% Fill in the following space with the manuscript text.
%
% A number of LaTeX commands may be invoked in this space, e.g.
%
% \section{} : to insert a new section title
% \label{}   : to label the numbered section for use in \ref{}
% \cite{}    : to add a reference using the label in \bibitem{}
% 
% A number of LaTeX environments may be used, e.g. 
% \begin{equation}
%     An equation inserted here will be automatically numbered
% \end{equation}  
%
% Please refer to other LaTeX documentation for help on using these
% environments.

%%Insert the main text of the manuscript here.

Strongly correlated electron systems display usual non-Fermi-liquid
behavior in the vicinity of zero-temperature phase transitions
\cite{Hertz76,Millis93}. The nature of these so-called quantum
critical points is still not well understood in real materials.
The two-channel Kondo effect (2CKE) is a prototype for such a
quantum critical point in a quantum impurity system. It occurs
when a spin-$\frac{1}{2}$ local moment is coupled
antiferromagnetically with equal strength to two independent
conduction-electron channels that overscreen its moment~\cite{CZ98}.
Its fixed point governs the transition between two distinct Fermi
liquids, which are adiabatically connected at finite temperature.

\begin{figure}[htbp]
  \centering
  \includegraphics[width=50mm]{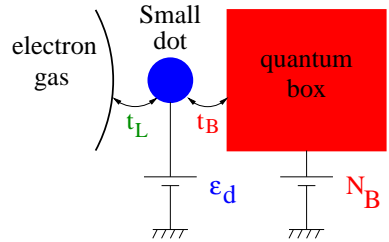}

  \caption{A sketch of an ultra-small grain coupled to a lead
           and to a quantum box.}
  \label{fig:model}
\end{figure}

The remarkable capabilities of detailed sample engineering and the
direct control of the microscopic model parameters have turned
quantum-dot devices into an important tool for investigating
fundamental questions such as the 2CKE. We explore a
double-dot device, comprised of a quantum box (i.e., a large
dot) indirectly coupled to a lead via an ultrasmall quantum
dot~\cite{LSA03}, see Fig.\ref{fig:model}. The Coulomb blockade
on the quantum box suppresses charge fluctuations at temperature
below the charging energy $E_C$ of the box, dynamically
generating two independent screening channels for the spin
on the ultrasmall quantum dot~\cite{OGG03}. The latter dot
is modeled by a single energy level $\epsilon_d$ and an
on-site Coulomb energy $U$, embedded between a metallic lead
and a quantum box. The quantum box is characterized by a dense
set of single-particle levels. Denoting the creation of an
electron with spin projection $\sigma$ on the dot by
$d_{\sigma}^{\dagger}$, the corresponding Hamiltonian reads
\begin{eqnarray}
&& { H} = \sum_{\alpha=L,B} \sum_{k,\sigma}
                     \epsilon_{\alpha k}
                     c^{\dagger}_{\alpha k\sigma}c_{\alpha k\sigma}
           + \; E_C \left ( \hat{n}_B - N_B \right )^2
\label{initial_Hamiltonian} \\
&& +\
      \epsilon_d \sum_{\sigma} d^{\dagger}_{\sigma} d_{\sigma}
           + U \hat{n}_{d \uparrow} \hat{n}_{d \downarrow}
      + \sum_{\alpha, k, \sigma} t_{\alpha}
      \left\{
            c^{\dagger}_{\alpha k\sigma}d_{\sigma} + {\rm H.c.}
      \right\} ,
\nonumber
\end{eqnarray}
where $c^{\dagger}_{L k \sigma}$ ($c^{\dagger}_{B k \sigma}$)
creates a lead (box) electron with momentum $k$ and spin
projection $\sigma$, $t_L$ ($t_B$) is the tunneling matrix
element between the quantum dot and the lead (box), and
$\epsilon_{L k}$ ($\epsilon_{B k}$) are the single-particle
levels in the lead (box). The excess number of electrons
inside the box, $\hat{n}_B$, is controlled by the
dimensionless gate voltage, $N_B$. Taking the single-particle
levels in the lead and box to have a common rectangular density
of states, $\rho(\epsilon) = \rho \theta ( D - |\epsilon|)$,
we define the energy scale $\Delta=\pi \rho t_L^2$ and use
it as our unit of energy.

% mapping
The Hamiltonian of Eq.~(\ref{initial_Hamiltonian}) is solved
using a recent adaptation of Wilson's NRG~\cite{Wilson75} to
the Coulomb blockade~\cite{LSA03_FR}. The main complication
with applying the NRG to this system stems from the long-range
interactions in energy space induced by the charging energy
$E_C$. We circumvent this problem by mapping the original model
onto an equivalent Hamiltonian, describing two noninteracting
bands coupled to a complex impurity~\cite{LSA03_FR}. The
resulting quantum impurity problem is then accurately solved
using the NRG.

\begin{figure}[tbp]
  \centering
  \includegraphics[width=75mm]{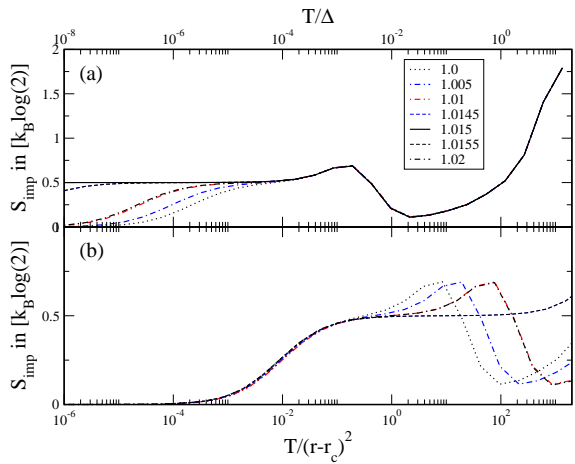}
  \caption{(a) The dot entropy
    $S_{\rm imp} = S_{full}-S_{free}$ vs temperature,
    for different values of $r = t_B/t_L$. (b) $S_{\rm imp}$
    replotted vs the rescaled temperature $T/(r-r_c)^2$.
    Parameters: $D=10$,
    $N_B=0$, $\epsilon_d = -1$, $U=2$, $E_C=0.01$, $N_s=1200$,
    and $\Lambda=5$.
  }
  \label{fig:entropy}
\end{figure}

{\em Results:} 
The impurity contribution to the entropy, $S_{\rm imp}$,
is defined as the difference between the entropy of
(\ref{initial_Hamiltonian}) with and without the
ultrasmall dot present. Figure~\ref{fig:entropy} (a)
shows $S_{\rm imp}(T)$ for different ratios $r = t_B/t_L$
near the quantum critical point (QCP) $r_c = 1.015$.
The dot parameters were chosen such that the crossover
temperature to the single-channel strong-coupling fixed
point, $T_s$, is larger than the charging energy $E_C$,
while the box is tuned to the plateau regime, $N_B = 0$.
With decreasing temperature, $S_{\rm imp}$ decreases rather
rapidly from the free-orbital value of $S_{\rm imp}=\log(4)$
towards the strong-coupling fixed-point value of
$S_{\rm imp} \to 0$. However, since the strong-coupling
fixed point is unstable with respect to $E_C$, the
entropy $S_{\rm imp}$ increases again as $T$ crosses
$E_C$~\cite{FR03}. The relevant operators associated
with $E_C$ have the effect of both inducing an effective
local moment, and suppressing the scattering between the
lead and the quantum box. At the QCP, the resulting Kondo
couplings between the effective moment and the two electron
gases are of equal strength. Hence the Hamiltonian flows to
the two-channel fixed point, characterized by a residual
entropy of $\frac{1}{2}\log(2)$. The two-channel
fixed point is unstable against a
channel-symmetry-breaking field, induced by a ratio
$r$ that deviates from $r_c$. Plot (b) in
Fig.~\ref{fig:entropy} shows the scaling behavior of
$S_{\rm imp}$ as a function of the scaling variable
$T/(r - r_c)^2$, using the data of (a). We report
perfect scaling for $T \ll E_C$, while the crossover
regime at intermediate and high temperatures is
governed by $E_C$ and $T_{s}$. Note that the overall
behavior of $S_{\rm imp}$ resembles the large-$N$
limit of Ref.~\cite{FR03}.

\begin{figure}[tbp]
  \centering
  \includegraphics[width=70mm]{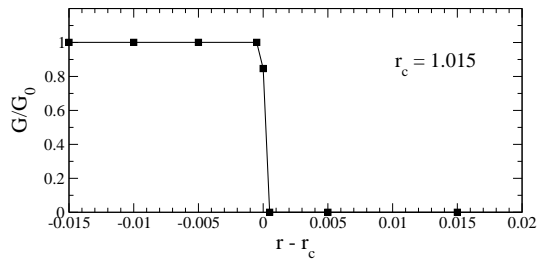}
  \caption{$T=0$ conductance as function of $r = t_B/t_L$,
	   for the same parameters as in Fig.~\ref{fig:entropy}.
	   Here $G_0$ is optimal conductance.
   }
  \label{fig:conductance}
\end{figure}

The different nature of the two Fermi liquids on either
side of $r_c$ is illustrated by the conductance in
Fig.~\ref{fig:conductance}. Now we drive a current
through the ultrasmall dot using two noninteracting
leads~\cite{OGG03}. For $r < r_c$, the effective
coupling to the quantum box scales to zero, rendering
the conductance perfect at $T=0$. For $r > r_c$, the
coupling to the leads scales to zero and the conductance
vanishes. Qualitatively, we find the same behavior when
the gate voltage $N_B$ is varied; details will be
published elsewhere.

E.L. and A.S. were supported in part by the Centers of
Excellence Program of the Israel Science Foundation.
F.B.A. acknowledges funding of the NIC, Forschungszentrum
J\"ulich, under project no. HHB00.

%----------------------------------------------------------------------
% Reference section
%
% List each reference with a separate \bibitem{} command.  The
% argument contains the label that is used in the \cite{} command
% in the main text
%
% e.g.
%
%    This follows our pioneering work on TdB2\cite{TdB2}.
%
% \bibitem{TdB2}
% J. Doe, J. Doe, and J. Dupont, J. Irrep. Res. 10 (2000) 1000.

%----------------------------------------------------------------------
% Figures and Tables
%
% Insert figures and tables at the end of the document unless you
% are familiar with the LaTeX positional options.
%
% \begin{figure}
%     \centering
%     \includegraphics{filename.eps}
%     \caption{Insert figure caption here} 
% \end{figure}  
%
% \begin{table}
%     \centering
%     \begin{tabular}
%     Insert table here
%     \end{tabular}
%     \caption{Insert table caption here}
% \end{table}  
%
% Please refer to other LaTeX documentation for help on using these
% environments.

%----------------------------------------------------------------------
% Terminate document

\end{document}